\documentclass{article}
\usepackage{graphicx} 
\usepackage{float} 
\usepackage{amsmath}
\usepackage[a4paper, total={6in, 8in}]{geometry}
\usepackage{cite}
\usepackage{hyperref}
\usepackage{soul}
\usepackage{xcolor}
\sethlcolor{yellow} 
\title{Grape expectations: direct thermographic imaging of electric fields in  microwave Mie resonators}

\author{Kallan K. Ronholm,$^{\dagger}$ Yuchen Song,$^{\dagger}$ Aaron D. Slepkov$^{\ast}$\\
\\
\normalsize{Department of Physics $\&$ Astronomy, Trent University,}\\
\normalsize{Peterborough, Ontario, Canada}\\
\normalsize{$^\ast$To whom correspondence should be addressed; E-mail:  aaronslepkov@trentu.ca} \\
\normalsize{$^\dagger$ These authors contributed equally to this work.}
}

\date{}
\begin{document}

\maketitle
\begin{abstract}
     
    Internal electromagnetic fields determine the behavior of dielectric nano-resonators, yet they remain exceptionally difficult to measure directly because subwavelength scales place them beyond conventional optical measurement, and any probe capable of local sensing perturbs the field it seeks to measure. We circumvent this limitation using water's high refractive index at microwave frequencies, which lets centimeter-scale aqueous spheres---grapes---serve as directly measurable, scale-invariant analogs of high-index nanophotonic resonators. Using polarization-controlled free-space excitation and infrared thermography, we show that internal heating patterns in grapes closely track simulated electric-, but not magnetic-, field distributions for both the fundamental magnetic- and electric-dipole Mie resonances. Extending this approach to grape dimers, we demonstrate that the massively sub-wavelength inter-object hotspot associated with microwave-induced sparking forms only under axial polarization, thus providing both a clue to its origin and a means for its control.
        
\end{abstract}

\section{Introduction}

 All-dielectric resonators have become an important complement to plasmonics in nanophotonics \cite{Krasnok12,Albella14,Kuznetsov16,Genevet17,Baranov17,Koshelev19}. Where plasmonic structures concentrate fields at metal surfaces and dissipate energy through ohmic loss~\cite{Khurgin15}, high-index dielectric resonators admit fields into their interior in addition to concentrating near-surface evanescent fields \cite{Devilez15,Kuznetsov12,Kuznetsov22}. The internal fields associated with resonant dielectric modes are increasingly of interest in their own right, with applications that include nano-lasers \cite{Ouyang24,Tiguntseva20}, non-radiative anapole resonators \cite{Miroshnichenko15,Gongora17,Baryshnikova19}, and enhanced absorbers \cite{Liu2016,Wen22}. Yet, the great majority of work on dielectric Mie scatterers characterizes them through their scattered far-field response \cite{Kruk17,Babicheva24,Krasnok14,Kuznetsov12}. Their near-field and internal-field distributions remain understudied experimentally, largely because subwavelength resolution restricts their direct optical observation, and because any probe capable of local measurement inevitably perturbs the dielectric landscape it is meant to characterize. \\
 
Mie theory describes dielectric scattering in terms of the ratio of object size to wavelength, making the underlying physics scale invariant \cite{Bohren83,Mie08}. This scale invariance offers a practical workaround to the resolution problem: rather than shrinking the probe, one can enlarge the resonator and the corresponding wavelength together, moving the physics into a regime where direct field measurement is straightforward. A second obstacle facing both the study and the application of dielectric nanoresonators reinforces the case for this approach: high-index dielectric materials are scarce at optical frequencies \cite{Khurgin22}, and considerable effort is presently directed at engineering (meta)materials with effective indices approaching 10 \cite{Vishnuramkumar2026}.\\

Water offers a way around both obstacles at once \cite{Wen22}. Its refractive index of approximately 9 at gigahertz frequencies means that centimeter-scale aqueous objects form natural microwave analogs of high-index nanophotonic resonators \cite{Andryieuski15,Jacobsen21,Song25}, and their macroscopic dimensions make internal and near-field concentrations directly accessible to measurement. Kapitanova \emph{et al}. used this correspondence to map internal magnetic fields in cylindrical water resonators using small inductive current probes \cite{Kapitanova17}.\\

Here, we exploit water's moderate microwave absorption to visualize, via thermography, the internal electric field concentrations associated with the first two Mie resonances of grapes. This builds on earlier findings that closely spaced grape pair ``dimers'' spark inside a household microwave oven \cite{khattak19soft,shen23}, a phenomenon we linked to Mie resonances in the individual grapes \cite{Khattak19}. Those works, however, relied on the uncontrolled electromagnetic environment of the oven cavity. Subsequent nanodiamond sensing by Fawaz et al. provided further evidence for an axial field hotspot in grape dimers \cite{Fawaz24}, but direct confirmation of the specific Mie modes responsible has remained elusive. We address this gap using polarization-controlled, free-space excitation of grapes at the calculated sizes of their first two fundamental dielectric resonances---a magnetic dipole mode and an electric dipole mode---and show that the resulting internal temperature maps closely track the predicted electric field distributions \cite{Bohren83,Pascale19}. We then extend this approach to a resonant grape dimer and show that formation of the axial hotspot responsible for sparking can be controlled through the polarization of the incident field.

\section{Materials and Methods}



\subsection{Microwave irradiation}
 
 Grapes were purchased at a local supermarket, and were selected for roundness and size. The density of the grapes averaged 1.1 g/ml, which is consistent with typical estimates of 80\%--85\% water by weight \cite{Bigard2020}. Prior to microwave irradiation, each sample is bisected into two symmetric hemispheres and promptly reassembled to maintain spherical structure. The pre-cut samples are placed on 3D-printed holders made of microwave-transparent polylactic acid (PLA), designed to keep the spherical samples fixed in a set orientation.\\ 

Microwave radiation is generated using a 900 watt, 2.45 Ghz, household microwave oven, modified with a custom rectangular steel horn transecting the oven chamber and abutted to the internal waveguide port. This provides linearly-polarized quasi-plane free-space microwaves, with timing and power control provided by the oven keypad. The horn opens into a 60 cm $\times$ 60 cm $\times$ 60 cm sample chamber lined with 1"-thick Cuming MT-26 microwave-absorbing foam. Control over the relative orientation between the sample geometry and the incident polarization axis was determined by rotating the sample holder, the oven and horn, or both.\\

A Flir T-540 SLR-type infrared thermal camera was affixed to the top of the sample chamber and used to record Movies of sample surface temperature during irradiation. All samples were placed 30 cm below the camera port and 50 cm horizontally from the end of the antenna and irradiated on full power for 8 seconds. The 8-second irradiation time provided sufficient heating to establish discernible temperature distributions without thermalization obscuring the patterns. Immediately after irradiation, the halves were separated to expose the desired interior ``cut planes'' for thermal imaging. The orthogonal cut planes are presented as a form of thermal tomography that allows for a visual reconstruction of the interior 3D volume. Temporal inspection of heated grape sections suggests that internal temperature maps are qualitatively stable for approximately 15 seconds in 1.4-cm samples and over 30 seconds in 1.9-cm samples. Nonetheless, in order to capture the thermal pattern with as little post-irradiation thermalization as possible, all thermal images presented here are still frames (extracted from the Movies) obtained within five seconds after microwave exposure. \\

\subsection{Simulations}
COMSOL Multiphysics is employed to simulate the interaction between microwaves and aqueous spheres within a terminated free-space simulation domain. The simulation model comprises a 15$\times$15$\times$20 $\text{cm}^3$ virtual cube with linearly polarized plane wave at $\lambda_0$ = 12.24 cm and a field magnitude of  $\text{E}_0=1$ V/m, incident from one plane upon a dielectric scatterer positioned at the center of the simulation volume. A second-order scattering boundary condition is applied to the simulation box to minimize reflections. The complex relative permittivity of the samples is set to $\tilde{\epsilon}=80+i10$, representing the dielectric properties of water at room temperature \cite{Vollmer03}. The relative permittivity of the environment is set to 1, approximating that of air.

\subsection{Mie theory}

In water, the wavelength of the incident microwave radiation is scaled down by the refractive index, and is thus comparable to the geometric dimensions of the aqueous samples, enabling the excitation of morphology dependent resonances. These optical resonances can be analytically described using multipole expansion techniques based on Mie scattering theory \cite{Mie08}. In general, the amplitudes of the magnetic and electric multipole modes are characterized by Mie scattering coefficients, which in turn comprise cross-sections that are used to describe resonance spectra. The complex magnetic and electric scattering coefficients, $\tilde{b}_n$ and $\tilde{a}_n$, respectively, are given by

\begin{align}
&   \tilde{b}_n = \frac{ \tilde{m}\, \psi_n(s)\, \psi_n'(\tilde{m} s) - \psi_n(\tilde{m} s)\, \psi_n'(s)}{\tilde{m}\, \xi_n(s)\, \psi_n'(\tilde{m} s) - \psi_n(\tilde{m} s)\, \xi_n'(s)}
\intertext{and}
&   \tilde{a}_n = \frac{ \tilde{m}\, \psi_n(\tilde{m} s)\, \psi_n'(s) - \psi_n(s)\, \psi_n'(\tilde{m} s)}{\tilde{m}\, \psi_n(\tilde{m} s)\, \xi_n'(s) - \xi_n(s)\, \psi_n'(\tilde{m} s)},
\end{align}\\
where the integer $n$ denotes mode order; $n$=1 corresponding to a dipole mode, $n$=2 to a quadrupole mode, $n$=3 to an octupole, etc.; $\tilde{m}\equiv \tilde{n}_1/\tilde{n}_0$ is the relative refractive index between the medium ($\tilde{n}_1=8.9+i0.6$) and the surrounding ($\tilde{n}_0=1$); $s= \frac{2\pi r n_0}{ \lambda_0}$ is a size parameter that relates the incident wavelength, $\lambda_0$, to the object radius, $r$; and $\psi$ and $\xi$ are Riccati-Bessel functions\cite{Bohren83}.\\

For a dielectric sphere, the partial extinction, scattering, and absorption cross sections associated with the $n^\text{th}$-order magnetic mode can be expressed in terms of the magnetic scattering coefficient, $b_n$~\cite{Kerker69}:\\
\begin{align}
&	\sigma_\text{ext}^{(b,n)} = \frac{(2n+1){\lambda_0}^2}{2\pi} \text{Re} \, [\tilde{b}_n] , \label{eq:Co}\\
&	\sigma_\text{sca}^{(b,n)} = \frac{(2n+1){\lambda_0}^2}{2\pi} |\tilde{b}_n|^2, \\ 
&	\sigma_\text{abs}^{(\text{Mag},n)} = \sigma_\text{ext}^{(b,n)}  - \sigma_\text{sca}^{(b,n)}.	\label{eq:Ca}
\end{align}\\
The expressions for the corresponding electric modes ($\sigma_\text{ext}^{(a,n)}$, $\sigma_\text{sca}^{(a,n)}$, and $\sigma_\text{abs}^{(\text{Elec},n)}$) are analogous and can be obtained by substituting the electric scattering coefficient $a_n$ for the magnetic scattering coefficient $b_n$ in Eqns.~\ref{eq:Co}--\ref{eq:Ca}. Ultimately, the absorption cross section quantifies the strength of the interaction between the incident microwaves and the aqueous sample. The total absorption cross section is obtained by summing the contributions from all electric and magnetic multipolar modes. Consequently, the absorption cross sections provide direct insight into the modal resonance behavior, enabling the identification and characterization of the underlying multipolar resonances.

\section{Results}
    


\begin{figure}[H]
        \centering
        \includegraphics[width=0.75\linewidth]{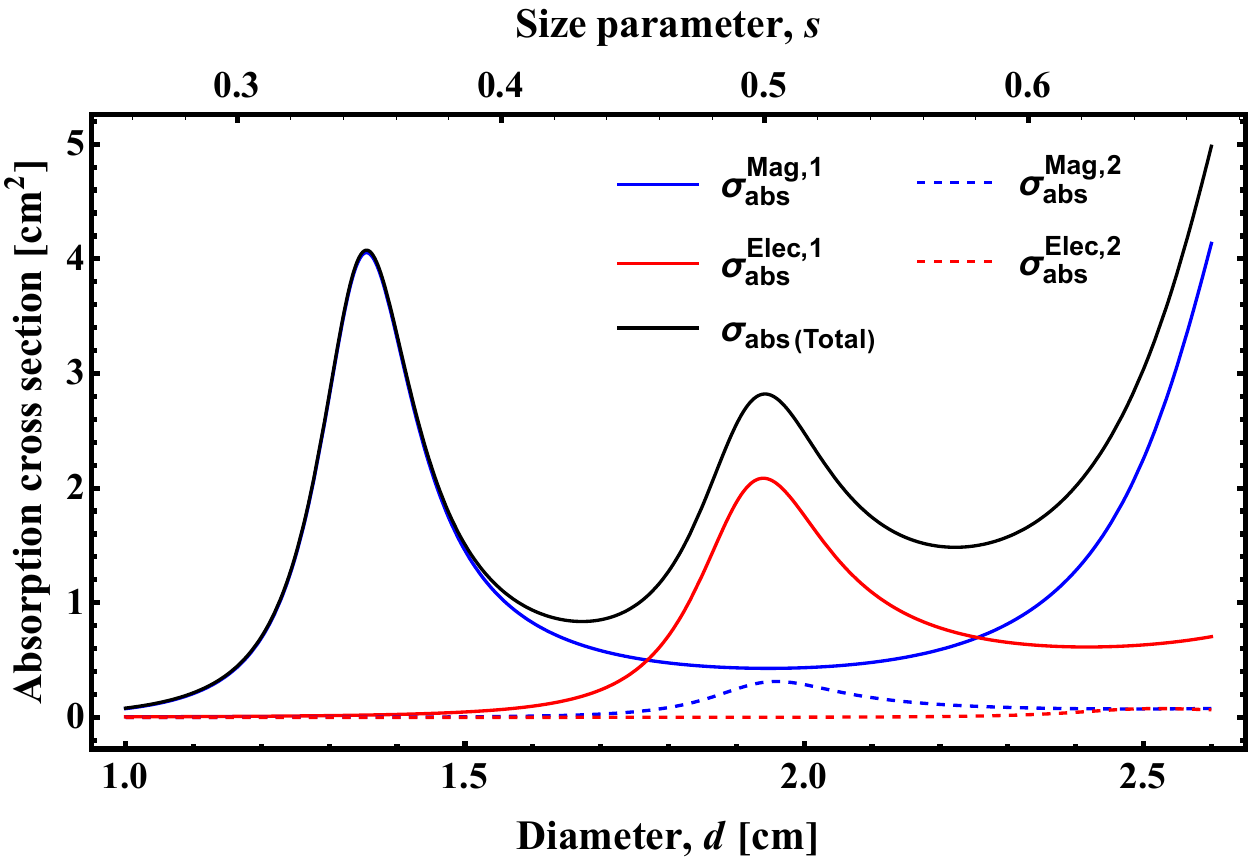}
        \caption{Calculated size-dependent absorption cross-sections for water spheres of diameter 1 to 2.5 cm.  The moderate absorption of water results in broadened and overlapped resonances. The solid curves denote the internal cross section for a magnetic dipole mode (MD, blue) and electric dipole mode (ED, red). Blue and red dashed lines represent the magnetic quadrupole (MQ) and electric quadrupole (EQ) modes, respectively. The total absorption cross section (black) is the sum of the electric and magnetic absorption cross sections. The first resonance is a well-isolated MD mode at 1.36 cm, whereas the next mode, at 1.92 cm, combines a dominant ED mode with a weak MQ mode and the tail of the MD resonance.}
        \label{fig:Excitation modes}
    \end{figure}

Whereas in highly transparent dielectrics the first few resonances are expected to be sharp, intense, and well-isolated, the absorption of microwaves in water broadens and weakens the underlying resonances in aqueous objects. The absorption spectrum of 1--2.5-cm aqueous spheres, as estimated via the total Mie absorption cross section
, is presented in Fig. \ref{fig:Excitation modes}. As can be seen in the figure, the first resonance---that is, the resonance excitable in the smallest object---is expected to be a pure magnetic dipole (MD) mode at a diameter of 1.36 cm. The second excitation, peaking at a diameter of 1.92 cm, is dominated by an electric diople (ED) mode, but combines contributions from a weak magnetic quadrupole (MQ) mode and the broadened tail of the primary MD mode. Because of the lack of purity of this resonance, we refer to this an ``electric-dipole-like'' mode. The next two resonances are expected at 2.68 cm and 3.30 cm, and are dominated by the second MD and ED resonances, respectively, with minor contributions from other multipoles. For experimental validation, we focus this work on the first two resonant sizes using grapes as representative aqueous spheres. Although not reported in detail here, all of the phenomena presented for grapes were independently confirmed in hydrogel ``water-bead'' analogs, where only minor cosmetic differences were observed.\\

\begin{figure}[H]
    \centering
    \includegraphics[width=0.9\linewidth]{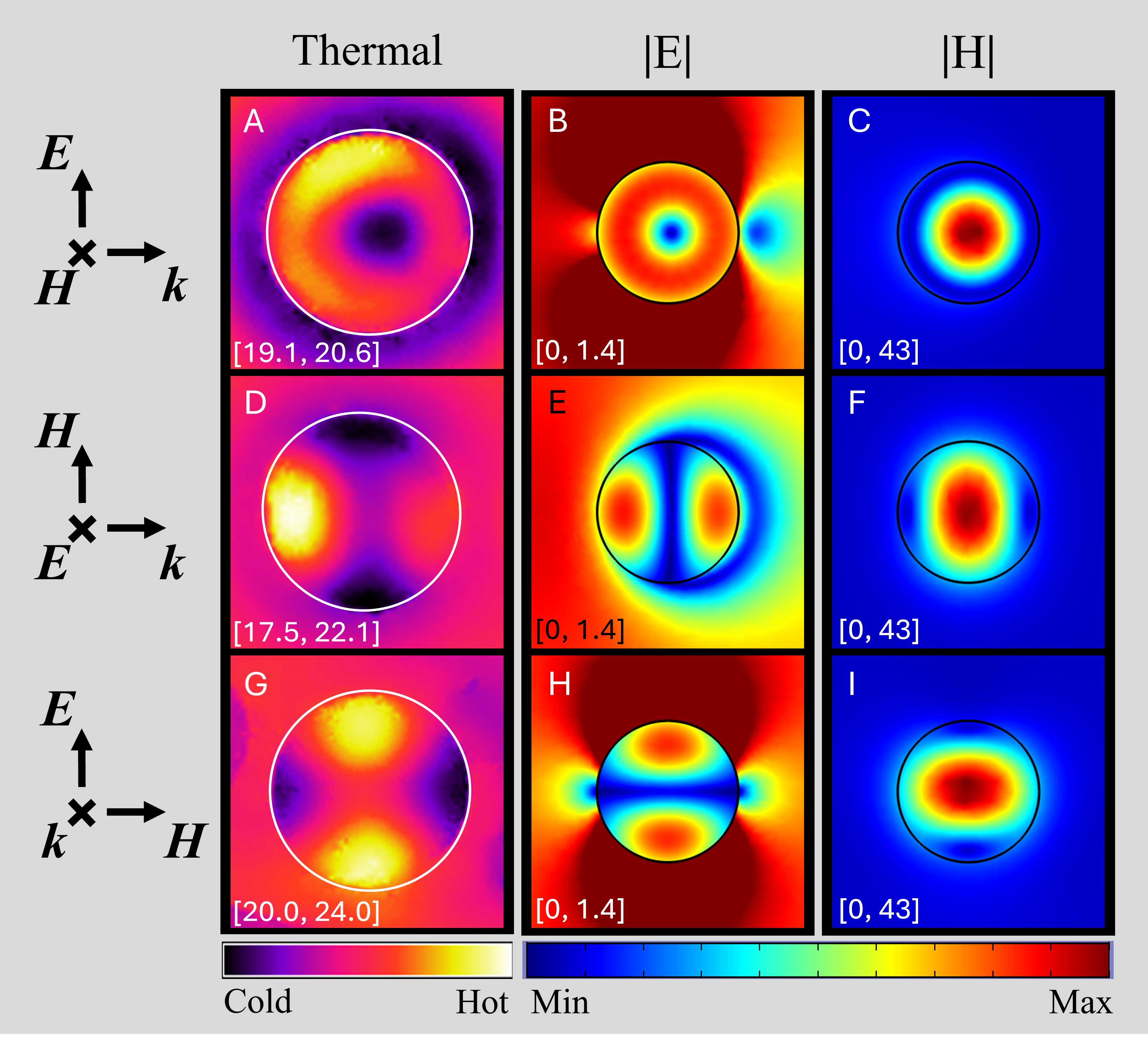}
    \caption{
     Comparison of experimental temperature maps along three cut planes in irradiated 1.36-cm-diameter grapes with simulated electric and magnetic field distributions in aqueous spheres at the same primary magnetic dipole resonance size. The grapes are irradiated for 8 seconds by a linearly polarized plane wave incident as labeled on the far left. Post-irradiation thermal images, panels (A), (D), and (G) are stills from Supplementary Movies S1, S2, and S3, respectively, with the values listed in the bottom left representing the minimum and maximum temperatures (in degrees Celsius) within each panel. Center panels (B), (E), and (H) present the simulated electric-field norms at the three cut planes, with inset values representing the minimum and maximum intra-panel values in V/m. Panels (C), (F), and (I) present the simulated magnetic field norms at the three cut planes, with inset values representing the minimum and maximum intra-panel values in mA/m. Simulation colourmap scales are min-max normalized to internal object values. Simulations include external scattered fields in the vicinity of the surface. Residing outside of the absorbing object, these external fields do not contribute to object heating and are therefore not detectable through thermal imaging. The temperature maps are qualitatively similar to the the simulated electric field maps, and distinct from the simulated magnetic field maps, establishing thermal imaging as an experimental tool for visualizing electric field concentrations in absorptive resonators.}
    \label{fig:MD_Monomer}
\end{figure}

In previous studies \cite{Khattak19,Song22}, radially-symmetric internal hotspots provided only circumstantial evidence for resonant heating that clearly deviates from a simple model of ``outside-in'' skin-depth-type heating. However, without the ability to control the incident polarization, the temperature map could neither be tied to a specific resonant mode, nor to a physical absorption mechanism. Because, for a given fundamental resonance, the electric and magnetic field distributions are expected to be distinct and orientationally-tied to the excitation geometry, the internal temperature map of a resonantly heated sample can confirm both the character of the resonance and the fact that heating occurs exclusively via electric-field absorption. This can be readily seen in the thermal images of pre-cut grapes. As shown in Fig. \ref{fig:MD_Monomer} for irradiated 1.36-cm grapes, the internal temperature distribution closely matches the simulated electric field pattern; the well-defined toroidal electric field distribution of a magnetic dipole mode. Conversely, the simulated magnetic field distribution---a columnar concentration aligned with the incident magnetic field---shows no discernible contribution to the temperature map. The agreement between the thermal and electric distributions is consistent across all three equatorial cut planes, indicating that heating is governed overwhelmingly by the electric field rather than by the magnetic field. Because water is a non-magnetic medium with a relative permeability approaching unity, the magnetic field of the incident wave does not directly couple to the dielectric via absorption, and thus does not lead to heating.\\

\begin{figure}[H]
        \centering
        \includegraphics[width=0.9\linewidth]{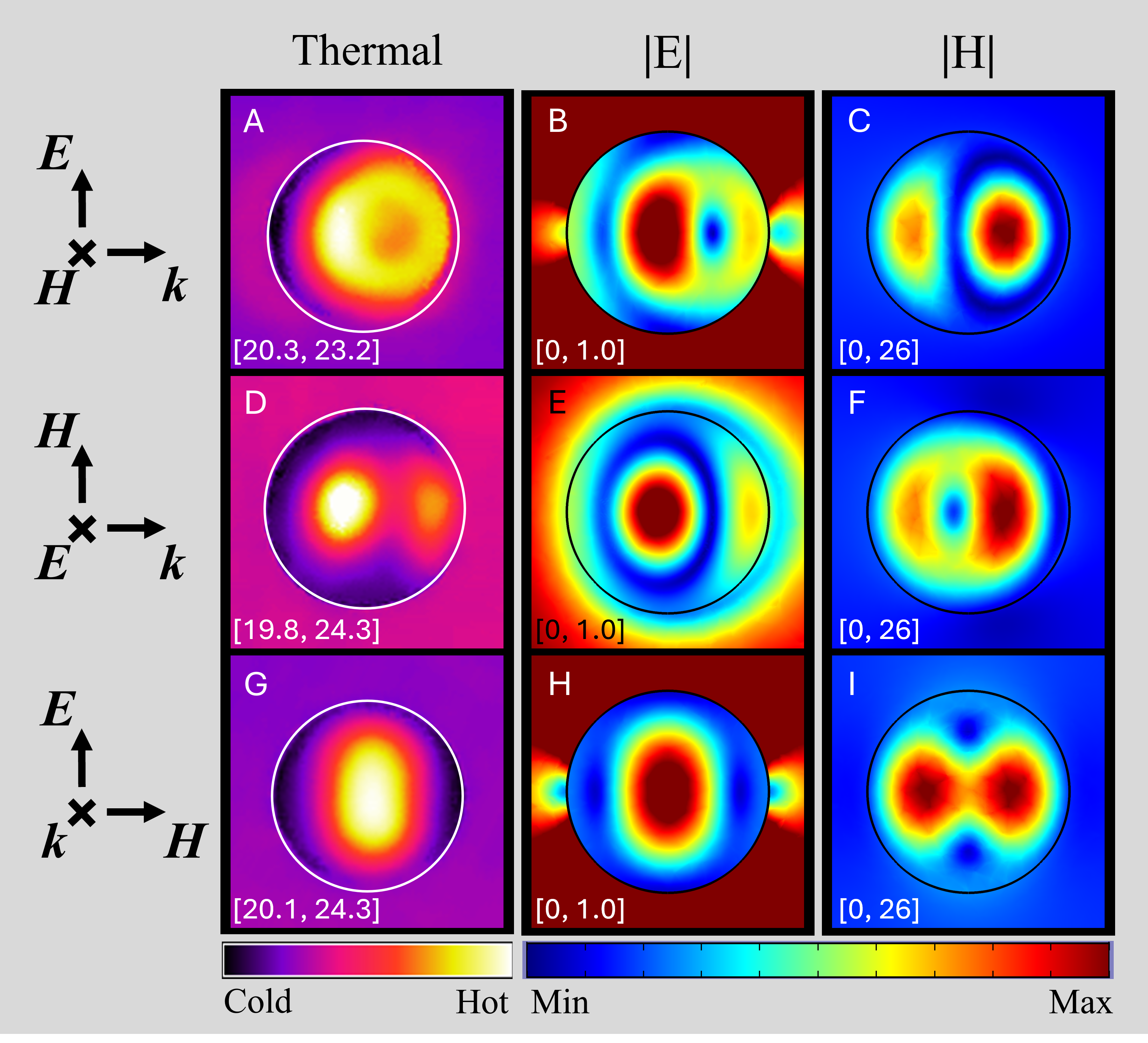}
          \caption{Comparison of experimental temperature maps along three cut planes in irradiated 1.95-cm-diameter grapes with simulated electric and magnetic field distributions in aqueous spheres at the same size. The grapes are irradiated for 8 seconds by a linearly polarized plane wave incident as labeled on the far left. Post-irradiation thermal images, panels (A), (D), and (G) are stills from Supplementary Movies S4, S5, and S6, respectively, with the values listed in the bottom left representing the minimum and maximum temperatures (in degrees Celsius) within each panel. Center panels (B), (E), and (H) present the simulated electric-field norms at the three cut planes, with inset values representing the minimum and maximum intra-panel values in V/m. Panels (C), (F), and (I) present the simulated magnetic field norms at the three cut planes, with inset values representing the minimum and maximum intra-panel values in mA/m. Simulation colourmap scales are min-max normalized to internal object values. Simulations include external scattered fields in the vicinity of the surface. Residing outside of the absorbing object, these external fields do not contribute to heating and are therefore not detectable through thermal imaging. The temperature maps are qualitatively similar to  the simulated electric field maps, and distinct from the simulated magnetic field maps, establishing thermal imaging as an experimental tool for visualizing electric field concentrations in absorptive resonators.}
        \label{fig:ED_Monomer}
    \end{figure}

Following the fundamental magnetic dipole resonance is an electric dipole resonance, for which the field distributions are swapped: The electric field is arranged as a circular column aligned with the incident polarization, and the magnetic field encircles it as a toroid. As established in Fig. \ref{fig:Excitation modes}, this excitation is expected to peak in a 1.95-cm-diameter sample and share some character with coexisting magnetic dipole and quadrupole modes. This can be readily seen in the thermal images of the larger grapes. As shown in Fig. \ref{fig:ED_Monomer} for irradiated 1.95-cm grapes, the internal temperature distribution closely matches the simulated electric field patterns, with no discernible contribution from the magnetic field distribution. As expected, the field patterns are dominated by magnetic dipole character, with some additional higher-spatial-frequency features arising from co-excitation of a weak quadrupolar mode. Crucially, the temperature maps of the two grape sizes are completely different, thereby allaying prior concerns that all previously-published thermal images in microwave-irradiated aqueous samples displayed similar radially-symmetric central hotspots irrespective of object diameter \cite{Khattak19,Song22}. \\

Another key difference between the modes in 1.36-cm and 1.95-cm grapes is a noticeable reduction in symmetry along the direction of incidence in the larger resonators. We attribute this asymmetry, which is most readily observable in Fig. \ref{fig:ED_Monomer} panels (A)-(F), to interference among the various modes coexisting at this diameter. This asymmetry is present in both the experimental temperature maps, and in the simulated electric and magnetic field distributions. In contrast, the minute asymmetry visible in the thermal images of the smaller, magnetic-dipole-resonant grapes has no counterpart in the simulations, and is most likely due to skin-depth absorption arising from the grapes' deviation from perfect orbs of water.

The first study to tie optical microwave resonances to the phenomenon of sparking in grapes hypothesized that an intense electromagnetic ``axial hotspot'' arises between two proximal grape resonators, which then initiates a plasma via field-ionization of sodium and/or potassium \cite{Khattak19}. However, because those experiments relied on microwave excitation with omnidirectional incidence and uncontrolled polarization, evidence tying the axial hotspot to the coupling of specific optical modes was relatively unconvincing. Furthermore, Lin \emph{et al.} \cite{Lin21} used parallel-plate capacitors to initiate sparking in grape dimers at 27 MHz---a frequency too low to excite optical resonances---and found that sparking only took place when the electric field was aligned with the axis of the dimer. Such orientationally-dependent excitation was presented as evidence that the axial hotpot (and associated sparking) arises from macroscopic electrical polarization charges, rather than from a photonic effect \cite{Bansal2022}. Nonetheless, we find that by controlling the directions of incidence and polarization, the axial hotspot can be selectively induced in free space at 2.5 GHz (i.e., optically). Furthermore, our demonstration of the coexistence of the axial hotspot and the internal dielectric resonance of the individual objects comprising the dimer suggests that the hotspot arises from, or is strengthened by, near-field mode coupling of the individual resonances. \\

\begin{figure}[H]
        \centering
        \includegraphics[width=0.9\linewidth]{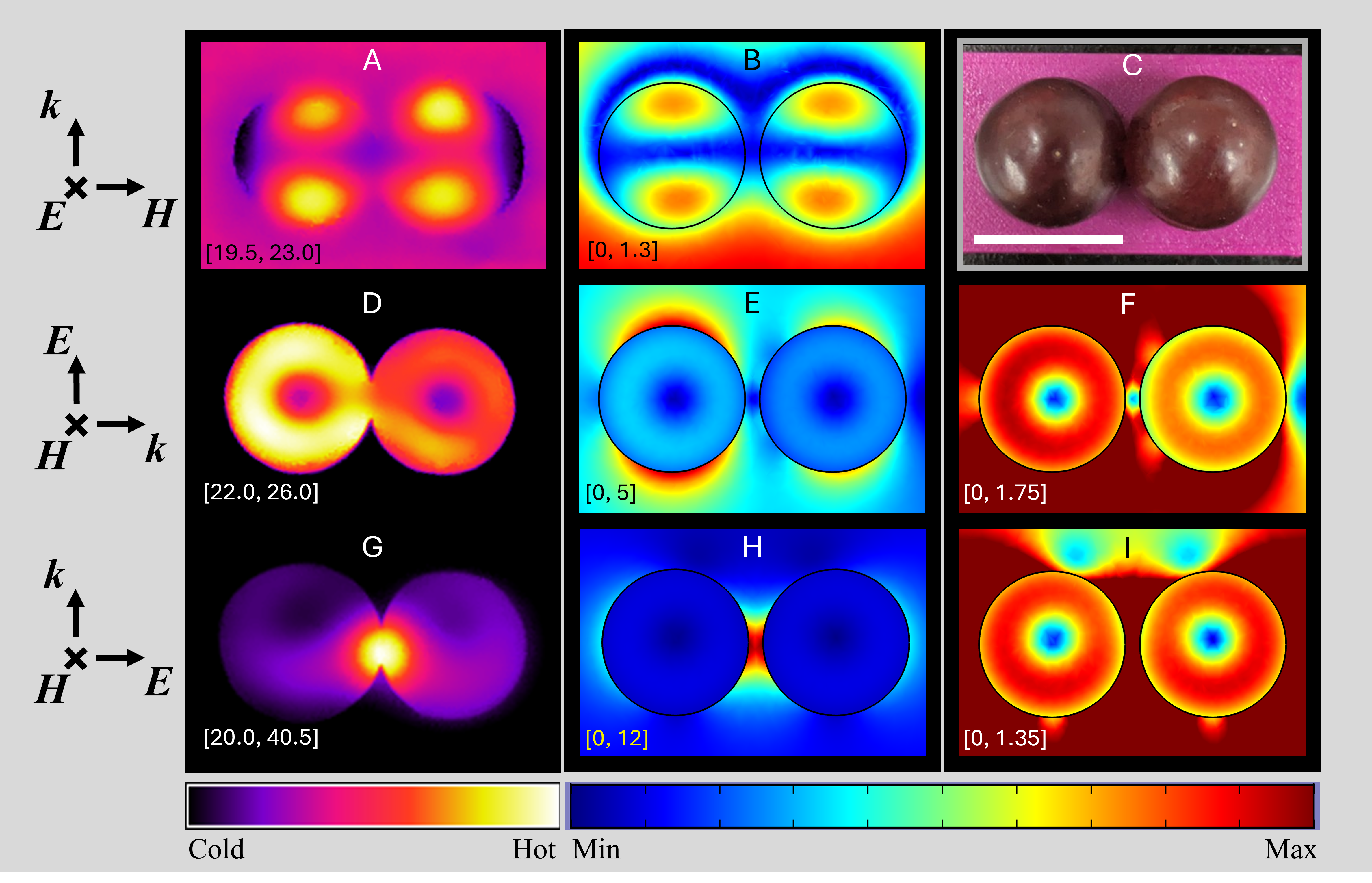}
        \caption{Comparison of experimental temperature maps along three cut planes in irradiated 1.36-cm-diameter grape dimers with the associated simulated electric field distributions in aqueous dimers of the same size. The far left of the figure describes the polarization and propagation of the incident microwaves. Post-irradiation thermal images, panels (A), (D), and (G) are stills from Supplementary Movies S7, S8, and S9, respectively, with the values listed in the bottom left representing the minimum and maximum temperatures (in degrees Celsius) within each panel. (Center: B, E, H) Corresponding simulated electric density maps, with the heat map scaled to global max. Values in the lower left are min and max, in volts per meter. (C) Is an in-situ photo of the pre-cut, pre-irradiated grape dimer (scale bar=1.36 cm). (F, H) are the simulated electric field densities of panels (E) and (H), but with the heatmap re-normalized to internal sample maximum. The axial hotspot in (G) is notably hotter than the maximum temperatures in both (A) and (D)}
        \label{fig:MD_Dimer}
    \end{figure}

Fig. \ref{fig:MD_Dimer} presents a comparison of the experimental temperature map and the corresponding simulations of the electric field distribution for a dimer of grapes, each sized at the fundamental magnetic dipole resonance (1.36-cm diameter). The dimer axis breaks the spherical symmetry of the monomers, providing three natural excitation geometries of interest based on the directions of propagation and polarization: two orientations in which the polarization is \emph{transverse} to the dimer axis and one orientation in which the polarization is \emph{axial}. As seen in Fig. \ref{fig:MD_Dimer}, simulations of electric field concentrations in the near field of the dimer suggest that for all three excitation geometries the magnetic dipole mode will be excited in each of the spheres comprising the dimer, with field maps matching those of Fig. \ref{fig:MD_Monomer}. However, (only) when the electric field is polarized axially, an intense and highly sub-wavelength electric field hotspot is predicted between the dimers. These predictions are experimentally confirmed by the temperature maps shown in Fig. \ref{fig:MD_Dimer} (A, D, and G). Most notably, an intense hotspot is observed at the point of dimer contact exclusively for axially-polarized incidence (Fig. \ref{fig:MD_Dimer} (G)). This sub-wavelength hotspot dominates and thus somewhat obscures the internal toroidal temperature distribution of the co-existing MD resonance.

Whereas a small amount of thermal diffusion is likely taking place in the objects, both during heating and in the interim between excitation and thermal imaging, the internal features of the resonant temperature maps are relatively stationary and provide a reasonable representation of the real size of the electromagnetic field distributions. This, however, does not likely hold for the axial hotspot itself because this hotspot is mostly superficial (i.e. not in the bulk of the resonator) and thus diminishes rapidly via radiative cooling at the termination of microwave exposure. This fact is discernible from the video sequence associated with Fig. \ref{fig:MD_Dimer} (G) (see Supplementary Movie S9). Much of this axial hotspot comprises residual diffused heat from the surface into the bulk of the grapes. Thus, the true electric-field concentration within the axial hotspot is likely at least an order of magnitude smaller still, representing approximately $\lambda_0/{100}$ focusing, as originally estimated using thermally-activated paper \cite{Khattak19}. \\ 

A notable asymmetry is observable in the case of axial incidence (transverse polarization), wherein the object closer to the source is slightly hotter than the one behind it. This asymmetry can be found both in the simulated field maps and in the thermal image (see Fig. \ref{fig:MD_Dimer}). This suggests that the first object receives the full incident intensity, whereas the second object is somewhat in the shadow of the first, and receives a lower, absorption-attenuated intensity. In the two other incident conditions, both objects comprising the dimer receive the full incident illumination, and thus there is no inter-object temperature asymmetry. A nearly-indiscernible intra-object temperature asymmetry is, however, present between the front and back of the object in these arrangements, possibly arising from simple skin-depth absorption, from a superposition of the MD mode and a tiny contribution of the ED mode, or from experimental deviation of grapes from perfect orbs of water.\\

Because thermal imaging is strictly a surface observation method, visualization of internal temperature distributions requires post-irradiation sample manipulation. Thus, a primary limitation of the present methodology is that the internal field distributions cannot be observed directly in real time. It is conceivable however that computational methods could be combined with better knowledge of material properties such as thermal conductivity to allow for reconstruction of internal temperature distributions from surface thermography, without the need for physical sample sectioning \cite{Anthony16,Groz20}. Furthermore, while reliable and repeatable, the present experimental setup is relatively rudimentary and could be significantly improved by the addition of continuous microwave power and polarization control, improved thermal imaging, and a more anechoic environment.


\section{Conclusion}

In this work we established thermographic imaging as a direct, non-perturbative method for visualizing the internal electric field structure of absorptive dielectric Mie resonators; fundamental modes that have been predicted analytically for over a century but have remained largely inaccessible to direct experimental observation. We exploited water's high refractive index at microwave frequencies to scale up nanophotonic physics to a directly measurable size, and excited the fundamental magnetic- and electric-dipole-like resonances of grapes, showing that their internal temperature distributions, imaged immediately after irradiation, closely track the corresponding simulated electric (but not magnetic) field patterns. That this correspondence is readily apparent even in an imperfect, biologically variable medium such as a grape speaks to the robustness and ubiquity of the underlying phenomenon of microwave dielectric resonances and of the usefulness of visualizing internal electric-field concentrations via local absorptive heating.\\

This approach was subsequently extended to coupled resonators by examining the creation of an intense axial hotspot in a grape dimer of the smallest resonant size. Whereas prior investigation of microwave grape dimer sparking was limited to experiments carried out inside of chaotic microwave environments, the careful control of both microwave polarization and dimer orientation in this study allowed us to confirm that the axial hotspot is selectively and controllably induced only under axial polarization, and that it co-exists with the same fundamental dielectric resonance we visualize in isolated grapes. This result illustrates the diagnostic power of the technique: by visualizing the internal electric fields, we are able to show the concurrent excitation of standalone optical resonances and the axial hotspot, thereby adding evidence for the photonic nature of the phenomenon.\\

Despite focusing on cm-scale aqueous spheres at microwave frequencies, the underlying phenomena described here are applicable to all dielectric systems, including those at the nanoscale. Analogous systems and effects are well established in nanoscale dielectric and plasmonic particles\cite{Pascale19}, where direct experimental access to internal field distributions remains elusive. Because our method requires no exotic instrumentation, no cryogenic environment, and no inversion of far-field scattering data, we anticipate it will find immediate use as a low-cost, intuitive proxy for validating internal-field predictions in a range of settings: benchmarking simulations of anapole and other non-radiative dielectric modes prior to nanoscale fabrication; visualizing near-field hotspots relevant to nanolaser and nanoantenna design; characterizing local field concentration in engineered metasurfaces and metamaterial unit cells; and informing thermal management strategies in high-index photonic devices. More broadly, this approach offers a general pathway for macroscopic, tabletop studies of microscopic light-matter interactions, making textbook electromagnetic phenomena that have previously been accessible through calculation or simulation available for direct and intuitive experimental observation.

\bibliography{Citation}

@article{Krasnok14,
	author = {Krasnok, Alex and Simovski, Constantin and Belov, Pavel and Kivshar, Yuri},
	journal = {Nanoscale},
	pages = {7354-7361},
	title = {Superdirective dielectric nanoantennas},
	volume = {6},
	year = {2014}}

@article{Gongora17,
	author = {Totero Gongora, Juan Sebastian and Favraud, Gael and Fratalocchi, Andrea},
	journal = {Nanotechnology},
	pages = {104001},
	title = {Fundamental and high-order anapoles in all-dielectric metamaterials via Fano--Feshbach modes competition},
	volume = {28},
	year = {2017}}

@article{Ouyang24,
	author = {Y. Ouyang and H. Luan and Z. Zhao and W. Mao and R. Ma},
	journal = {Nature},
	pages = {287-293},
	title = {Singular dielectric nanolaser with atomic-scale field localization},
	volume = {632},
	year = {2024}}

@article{Khurgin15,
	author = {J. B. Khurgin},
	journal = {Faraday Discuss},
	pages = {109-122},
	title = {Ultimate limit of field confinement by surface plasmon polaritons},
	volume = {178},
	year = {2015}}

@article{Baranov17,
	author = {Baranov, Denis and Zuev, Dmitry and Lepeshov, Sergei and Kotov, Oleg and Krasnok, Alex and Evlyukhin, Andrey and Chichkov, Boris},
	journal = {Optica},
	pages = {814-825},
	title = {All-dielectric nanophotonics: the quest for better materials and fabrication techniques},
	volume = {4},
	year = {2017}}

@article{Kuznetsov22,
	author = {Kuznetsov, Aleksei and Can{\'o}s Valero, Adri{\`a} and Shamkhi, Hadi K. and Terekhov, Pavel and Ni, Xingjie and Bobrovs, Vjaceslavs and Rybin, Mikhail and Shalin, Alexander},
	journal = {Sci. Rep.},
	pages = {21904},
	title = {Special scattering regimes for conical all-dielectric nanoparticles},
	volume = {12},
	year = {2022}}

@article{Miroshnichenko15,
	author = {Miroshnichenko, Andrey and Evlyukhin, Andrey and Yu, Ye Feng and Bakker, Reuben and Chipouline, A. and Kuznetsov, Arseniy and Luk'yanchuk, Boris and Chichkov, Boris and Kivshar, Yuri},
	journal = {Nature communications},
	pages = {8069},
	title = {Nonradiating anapole modes in dielectric nanoparticles},
	volume = {6},
	year = {2015}}

@article{Genevet17,
	author = {Genevet, Patrice and Capasso, Federico and Aieta, Francesco and Khorasaninejad, Reza and Devlin, Robert},
	journal = {Optica},
	pages = {139-152},
	title = {Recent advances in planar optics: from plasmonic to dielectric metasurfaces},
	volume = {4},
	year = {2017}}

@article{Albella14,
	author = {P. Albella and R. {Alcaraz de la Osa} and F. Moreno and S. A. Maier},
	journal = {ACS Photonics},
	pages = {524-529},
	title = {Electric and magnetic field enhancement with ultralow heat radiation dielectric nanoantennas: considerations for surface-enhanced spectroscopies},
	volume = {1},
	year = {2014}}

@article{Kruk17,
	author = {S. Kruk and Y. Kivshar},
	journal = {ACS Photonics},
	pages = {2638-2649},
	title = {Functional meta-optics and nanophotonics governed by {Mie} resonances},
	volume = {4},
	year = {2017}}

@article{Kapitanova17,
	author = {P. Kapitanova and V. Ternovski and A. Miroshnichenko and N. Pavlov and P. Belov and Y. Kivshar and M. Tribelsky},
	journal = {Sci. Rep.},
	pages = {731},
	title = {Giant field enhancement in high-dielectric subwavelength particles},
	volume = {7},
	year = {2017}}

@article{Pascale19,
	author = {M. Pascale and G. Miano and R. Tricarico and C. Forestiere},
	journal = {Sci. Rep.},
	pages = {14524},
	title = {Full-wave electromagnetic modes and hybridization in nanoparticle dimers},
	volume = {9},
	year = {2019}}

@article{Andryieuski15,
	author = {A. Andyieuski and S. M. Kuznetsova and S. V. Zhukovsky and Y. S. Kivshar and A. V. Lavrinenko},
	journal = {Sci. Rep.},
	pages = {13535},
	title = {Water: Promising opportunities for tunable all-dielectric electromagnetic Metamaterials},
	volume = {5},
	year = {2015}}

@article{Babicheva24,
	author = {Viktoriia E. Babicheva and Andrey B. Evlyukhin},
	journal = {Adv. Opt. Photon.},
	pages = {539-658},
	title = {Mie-resonant metaphotonics},
	volume = {16},
	year = {2024}}

@book{Kerker69,
	author = {M. Kerker and E. M. Loebl},
	publisher = {Academic Press, New York},
	title = {The Scattering of Light and Other Electromagnetic Radiation},
	year = {1970}}

@article{Kuznetsov12,
	author = {A. I. Kuznetsov and A. E. Miroshnichenko and Y. H. Fu and J. Zhang and B. Luk'yanchuk},
	journal = {Sci. Rep.},
	pages = {492},
	title = {Magnetic light},
	volume = {2},
	year = {2012}}

@article{Vollmer03,
	author = {M. Vollmer},
	journal = {Phys. Educ.},
	pages = {74},
	title = {Physics of the microwave oven},
	volume = {39},
	year = {2003}}

@article{Kuznetsov16,
	author = {A. I. Kuznetsov and A. E. Miroshnichenko and M. L. Brongersma and Y. S. Kivshar and B. Luk'yanchuk},
	journal = {Science},
	pages = {aag2472},
	title = {Optically resonat dielectric nanostructures},
	volume = {354},
	year = {2016}}

@book{Bohren83,
	author = {C. F. Bohren and D. R. Huffman},
	publisher = {Wiley, New York},
	title = {Absorption and Scattering of Light by Small Particles},
	year = {1983}}

@article{Khattak19soft,
	author = {H. K. Khattak and S. R. Waitukaitis and A. D. Slepkov},
	journal = {Soft Matter},
	pages = {5804},
	title = {Microwave induced mechanical activation of hydrogel dimers},
	volume = {15},
	year = {2019}}

@article{Khattak19,
	author = {H. K. Khattak and P. Bianucci and A. D. Slepkov},
	journal = {Proc. Natl. Acad. Sci. USA},
	pages = {4000},
	title = {Linking plasma formation in grapes to microwave resonances of aqueous dimers},
	volume = {116},
	year = {2019}}

@article{Song22,
	author = {Y. Song and J. Shafe-Purcell and A. Slepkov},
	journal = {AIP. ADV},
	pages = {115216},
	title = {Linking microwave heating in aqueous spheres to morphology-dependent resonances},
	volume = {12},
	year = {2022}}

@article{Khurgin22,
	author = {J. B. Khurgin},
	journal = {ACS Photonics},
	pages = {743-751},
	title = {Expanding the photonic palette: exploring high index materials},
	volume = {9},
	year = {2022}}

@article{Krasnok12,
	author = {A. E. Krasnok and A. E. Miroshnichenko and P. A. Belov and Y. Kivshar},
	journal = {Opt. Express},
	pages = {20599},
	title = {All-dielectric optical nanoantennas},
	volume = {20},
	year = {2012}}

@article{Koshelev19,
	author = {K. Koshelev and G. Favraud and A. Bogdanov and Y. Kivshar and A. Fratalocchi},
	journal = {Nanophotonics},
	pages = {725-745},
	title = {Nonradiating photonics with resonant dielectric nanostructures},
	volume = {8},
	year = {2019}}

@article{Devilez15,
	author = {A. Devilez and X. Zambrana-Puyalto and B. Stout and N. Bondod},
	journal = {Phys. Rev. B},
	title = {Mimicking localized surface plasmons with dielectric particles},
	volume = {92},
	year = {2015}}

@article{Mie08,
	author = {G. Mie},
	journal = {Annalen Der Physik},
	pages = {377-445},
	title = {Beitr{\"a}ge zur Optik tr{\"u}ber Medien, speziell kolloidaler Metall{\"o}sungen},
	volume = {330},
	year = {1908}}

@article{Song25,
	author = {Y. Song and M. Hu and A. Slepkov},
	journal = {Phys. Rev. A},
	title = {Predicting microwave resonances in aqueous spheroids with a standing-wave approach},
	year = {2025}}

@article{Jacobsen21,
	author = {R. E. Jacobsen and S. Aeslanagic and A. V. Lavrinenko},
	journal = {Appl. Phys. Rev.},
	pages = {041304},
	title = {Water-based devices for advanced control of electromagnetic waves},
	volume = {8},
	year = {2021}}

@article{Lin21,
    author = {Lin, M. S. and Liu, L. C. and Barnett, L. R. and Tsai, Y. F. and Chu, K. R.},
    title = {On electromagnetic wave ignited sparks in aqueous dimers},
    journal = {Physics of Plasmas},
    volume = {28},
    number = {10},
    pages = {102102},
    year = {2021},
    issn = {1070-664X},
    doi = {10.1063/5.0062014},
    url = {https://doi.org/10.1063/5.0062014}}

@article{Shen23,
	author = {L. Shen and Q. Ran and X. Zhang},
	journal = {Applied Physics Letters},
	pages = {224101},
	title = {Inhibition effects of the applied dielectric on dimer-induced microwave plasma and focused hotspots},
	volume = {122},
	year = {2023}}

@article{Fawaz24,
	author = {Ali Fawaz and Sarath Raman Nair and Thomas Volz},
	journal = {Phys. Rev. Applied},
	pages = {064078},
	title = {Coupling nitrogen-vacancy center spins in diamond to a grape dimer},
	volume = {22},
	year = {2024}}

@article{Vishnuramkumar2026,
	author = {D. Vishnuramkumar and D. Singh},
	journal = {ACS Appl. Opt. Mater.},
	pages = {314-324},
	title = {High-Index Dielectric Metasurfaces: Unlocking Frontiers in Mid-Infrared Photonics},
	volume = {4},
	year = {2026}}

@article{Bigard2020,
    author = {Bigard, Antoine and Romieu, Charles and Sire, Yannick and Torregrosa, Laurent},
	journal = {Front. Plant Sci.},
	pages = {1175},
	title = {Vitis vinifera L. Diversity for Cations and Acidity Is Suitable for Breeding Fruits Coping With Climate Warming},
	volume = {11},
	year = {2020}}

@article{Anthony16,
author = {D. Anthony and D. Sarkar and A. Jain},
journal = {Sci. Rep.},
pages = {35886},
title = {Non-invasive, transient determination of the core temperature of a heat-generating olid body},
volume = {6},
year = {2016}
}

@article{Groz20,
author = {Groz, Marie-Marthe and Abisset-Chavanne, Emmanuelle and Meziane, A. and Sommier, Alain and Pradere, Christophe},
year = {2020},
pages = {1607},
title = {Bayesian Inference for 3D Volumetric Heat Sources Reconstruction from Surfacic IR Imaging},
volume = {10},
journal = {Applied Sciences}
}

@article{Baryshnikova19,
author = {Baryshnikova, Kseniia and Smirnova, Daria and Luk'yanchuk, Boris and Kivshar, Yuri},
year = {2019},
month = {01},
pages = {1801350},
title = {Optical Anapoles: Concepts and Applications},
volume = {7},
journal = {Advanced Optical Materials}
}

@article{Tiguntseva20,
author = {Tiguntseva, Ekaterina and Koshelev, Kirill and Furasova, Aleksandra and Tonkaev, Pavel and Mikhailovskii, Vladimir and Ushakova, Elena V. and Baranov, Denis G. and Shegai, Timur and Zakhidov, Anvar A. and Kivshar, Yuri and Makarov, Sergey V.},
title = {Room-Temperature Lasing from Mie-Resonant Nonplasmonic Nanoparticles},
journal = {ACS Nano},
volume = {14},
number = {7},
pages = {8149-8156},
year = {2020},
}

@article{Liu2016,
author = {Xiaoming Liu and Ke Bi and Bo Li and Qian Zhao and Ji Zhou},
journal = {Opt. Express},
number = {18},
pages = {20454--20460},
publisher = {Optica Publishing Group},
title = {Metamaterial perfect absorber based on artificial dielectric ``atoms''},
volume = {24},
year = {2016},
url = {https://opg.optica.org/oe/abstract.cfm?URI=oe-24-18-20454},
doi = {10.1364/OE.24.020454},
}

@article{Wen22,
author = {Jingda Wen and Qian Zhao and Ruiguang Peng and Haoyang Yao and Yuchang Qing and Jianbo Yin and Qiang Ren},
journal = {Opt. Mater. Express},
number = {4},
pages = {1461--1479},
title = {Progress in water-based metamaterial absorbers: a review},
volume = {12},
month = {Apr},
year = {2022},
url = {https://opg.optica.org/ome/abstract.cfm?URI=ome-12-4-1461},
doi = {10.1364/OME.455723},
}

@article{Bansal2022,
  author={Bansal, Rajeev},
  journal={IEEE Antennas Propag. Mag.}, 
  title={Wrath of Grapes Redux}, 
  year={2022},
  volume={64},
  number={1},
  pages={131-131},
  doi={10.1109/MAP.2021.3129693}}

\section*{Acknowledgments}
We thank Shima Nikkah Fini for preliminary work related to higher-mode dimer excitation; Miao Hu for early theoretical support; Hamza Khattak and Rodion Gordzevich for initial assembly of the microwave horn antenna; and Prof. Pablo Bianucci for fruitful discussion and moral support. 
\paragraph*{Funding:} This work was supported by the Natural Sciences and Engineering Research Council of Canada (NSERC), grant RGPIN-2024-03874.
\paragraph*{Author contributions:}
KR, YS, and ADS designed the study, discussed all findings, and wrote the manuscript. KR performed experiments under the guidance of ADS and YS, prepared the figures, and wrote the first manuscript draft. YS performed theoretical calculations and conducted simulations. ADS conceived and supervised the work.
\paragraph*{Competing interests:}
There are no competing interests to declare.
\paragraph*{Data and materials availability:}
COMSOL simulation outputs and code will be uploaded to a publicly-available repository. All other data is available in the main text or the supplementary materials.

\end{document}